\documentclass[12pt]{iopart}

\usepackage{graphicx}
\begin{document}

\title{Optimization in task--completion networks}

\author{L Dall'Asta$^1$, M Marsili$^1$ and P Pin$^{1,2}$}

\address{
\begin{itemize}
\item[\bf 1 \ ] Abdus Salam International Center for Theoretical Physics, Strada
  Costiera 11, 34014, Trieste, Italy
\item[\bf 2 \ ] Dipartimento di Scienze Economiche, Universit\`a Ca' Foscari di Venezia, Italy
\end{itemize}}

\ead{\mailto{dallasta@ictp.it}, \mailto{marsili@ictp.it}, \mailto{pin@unive.it}}

\begin{abstract}
We discuss the collective behavior of a network of individuals that receive, process and forward to each other tasks.
Given costs they store those tasks in buffers, choosing optimally the frequency at which to check and process the buffer.
The individual optimizing strategy of each node determines the aggregate behavior of the network.
We find that, under general assumptions, the whole system exhibits coexistence of equilibria and hysteresis.
\end{abstract}

\date{December 2007}
\ackn{M.M. and P.P. acknowledge support from EU-STREP project n. 516446 COMPLEX-MARKETS.}
\maketitle

\section{Introduction}

The study of the interaction between people processing tasks is at the base of the science of social networks \cite{M67}.
Recently the huge amounts of data coming from the analysis of e-mail networks,
web communities and social databases, have motivated the study of
statistical properties and related models of such human activities \cite{J04,B05,OB05,VODGKB06,BH07}.

A more complete analysis should take into account also the
`costs' associated to the execution and storage of tasks, that transform any individual
decision in a non-trivial economic optimization problem.
By costs we broadly intend any negative return deriving from the process.
In the case of execution it could be the effort and time spent, in the case of storage it could be the risk that the information or the payoff conveyed by the task become too old and hence useless.
In such cases the problem would be that of finding an optimal balance between those two different aspects.
Human activity is thus the result of a complex intertwined two-levels problem.
On the one hand, there is the individual problem
of  \emph{optimal timing} (or \emph{optimal stopping} \cite{PS06}) encountered by
rational agents in many economic situations, and object of investigations in operations research (as in optimal delivering \cite{MB97}), finance
(e.g. in the case of American options, see \cite{H89}), and macroeconomics (e.g. optimal retirement, as in \cite{CL81}).
On the other hand, we have the network queueing problem
resulting from the interactions between individuals, deeply studied in computer science \cite{GH98}
and recently in statistical physics as well \cite{KNGTR03,TRT07}.

The general problem can be formulated in the following way.
Let us consider a network of rational agents sending each others different tasks to be executed.
An agent gets positive payoff from processing a given task, but faces also some fixed
 and variable costs.
For this reason the agents can store tasks in queues and decide the optimal time intervals at which the
queues are checked and processed.
There is however a trade--off in delaying task processing, because there is always the risk, waiting too long, that tasks becomes too old to be worth being processed.

In principle, if tasks have different priorities (according to the models proposed in the literature \cite{B05,VODGKB06,BH07,C54,AW97}), every agent should have
different queues with different costs, each one containing tasks with similar priority.
The optimization problem associated with each queue produces different processing times,
and we expect that tasks' heterogeneity would be reflected in the heterogeneity of processing times as observed in real data \cite{OB05,HP04,DALRSB06,SKKHT06}.
Whereas a complete understanding of task-completion processes in human dynamics seems at present very
challenging, a first step in this direction is represented by understanding the relation between
individual optimization problems and the collective behavior.

In the present paper we study a version of this general model in which
individuals on a homogeneous network receive tasks from an external source and interact by processing
 and forwarding each other those tasks.
All tasks have the same priority and the individual problem consists in computing the optimal processing rate
that maximizes the individual utility.
Because of interaction, the individual strategies depend on the global behavior of the system, producing
 interesting collective phenomena.
We find analytically that, for some values of the parameters (i.e. external arrival rate of tasks, and costs), there are two possible stable equilibria for the collective optimal strategies: one in which nobody process tasks (trivial equilibrium), and one in which every individual
process tasks with the same non-zero rate.
Depending on the nature of costs, the transition between the two behaviors may not be continuous, and interesting hysteresis phenomena can be observed.

Understanding the effects of costs on the individual strategy is far from being of purely theoretical interest.
 Indeed, American e-mail providers AOL and Yahoo proposed in February 2006 a taxation on e-mails in
order to reduce the phenomenon of spamming.\footnote[6]{See e.g. \emph{The New York Times} from February 5 2006, online at: \\ http://www.nytimes.com/2006/02/05/technology/05AOL.html .}

The paper is organized in the following way. In Section \ref{individual} we focus on the individual behavior, characterizing the optimal responding strategy under the assumption of Poisson statistics for the arriving tasks.
We leave part of the detailed calculations, for this and the following sections, to \ref{computations}.
In Section \ref{interaction} we discuss the network process, and derive the stationary states under the assumption that single individuals respond optimally.
We assume, in order to obtain clear analytic results, that the network is homogeneous of fixed degree.
It is clear that the network structure will heavily depend on the specific environment (e.g. even for e--mails networks alone, different datasets have shown different topologies \cite{EMB02,GDDGA03}).
Section \ref{results} illustrates the main results about the collective behavior of the model, depending on the external parameters.
Section \ref{non_hom} generalizes to non--regular networks.
Conclusions and plans for future research are left to Section \ref{conc}.

\section{Individual behavior} \label{individual}

In this section we focus on the optimization of individual response strategies when the agents receive homogeneous tasks
(e.g. e-mails, jobs,\dots) to process and/or forward.
Tasks arrive from outside, in general both from an exogenous source and from other agents, but we ignore this distinction for the moment.
The only relevant assumption on individual behavior is that the agents have limited knowledge of system's dynamics, thus they always perceive the arriving
tasks as Poisson independent events. The arrival rate $\lambda$ is estimated monitoring the queue during the dynamics.

Arriving tasks are stored automatically in a queue (e.g. e-mail buffer), and the agent gets payoffs from processing tasks in her queue.
Each stored task has the same positive payoff $\pi >0$, but this payoff becomes $0$ at a Poisson rate $\mu$. Tasks with zero payoff are nor executed
neither forwarded.\footnote{If they are e--mails, the information they convey may become outdated. Generally, in the case of jobs or other opportunities, it could suddenly become too late to process them.}
Among the tasks which are stored in the buffer, we will call {\em survived} those that maintain the positive payoff.

Another important ingredient of real human dynamics is that individuals are not constantly monitoring their tasks list. For instance, people have to
connect to the Internet in order to check their web-mail window, and employees that have constant access to the web cannot spend their work-time
constantly monitoring their mail-box. Note that this cost can be arbitrarily small.
Therefore, in our model an agent is not constantly monitoring her buffer, and some fixed cost $C>0$ is paid whenever the queue is accessed.
Let us suppose that, opening the buffer, an agent finds $\ell$ new-arrived tasks, a non-negative cost $c<\pi$ is associated to check
 each of them.\footnote{The cost $c$ can be thought as the cost of reading an e-mail or just checking if a job is still worth being processed.}
As anticipated, the agent processes only the $n$ tasks that still give a positive payoff $\pi$.
Assuming a separable utility function, the overall net payoff is $n \pi - \ell c$.
Note that in this model, an agent does not execute the tasks bringing null utility;
this could be formalized with an additional positive cost $\kappa$ for processing tasks.
This cost would only re--scale $\pi$ to $\pi - \kappa$, so that, without loss of generality, we ignore this detail in the rest of the paper.

The individual optimization problem is the following: the agent has to choose a Poisson rate $\nu$ at which to check her buffer (and consequently process it).
Note that the condition $c < \pi$ prevents the trivial strategy, according to which it is always better not to check.
We normalize costs and payoffs, setting $\pi \equiv 1$, so that $c$ and $C$ are just ratios, with the property that $c < 1$.
Without loss of generality we also normalize the time setting $\mu \equiv 1$ so that our new $\lambda$ will be actually the ratio $\frac{\lambda}{\mu}$, and similarly for $\nu$.

The problem is to maximize the long run ($\tau \rightarrow \infty$) payoff per unit time, from the present time onwards:
\begin{eqnarray}
\max_{\nu} \left\{  \lim_{\tau \rightarrow \infty} \frac{1}{T(\nu)} \int_{\tau}^{\tau+T(\nu)} U(\nu, \lambda) \right\} & = &
\max_{\nu} \frac{E_{\nu, \lambda} [ U(\nu, \lambda) ]}{E_{\nu} [T(\nu)]} \nonumber \\
& = & \max_{\nu} \frac{E_{\nu, \lambda} [ n(\nu, \lambda) - \ell (\nu, \lambda) c - C] }{1 / \nu} ~. \label{opt1}
\end{eqnarray}
$E_{\nu}[T(\nu)]=\frac{1}{\nu}$ is the expected time between two checks of the queue, while $E_{\nu,\lambda}[\ell (\nu, \lambda)]$ and $E_{\nu,\lambda}[n(\nu, \lambda)]$ are respectively the expected total arrived and survived tasks in the buffer. All expectations are taken on the stochastic process with fixed $\lambda$, that is a parameter at the individual level, and $\nu$, the free quantity to tune the optimization process.
The optimization problem can also be derived from a Bellman equation, in the limit where the discount rate goes to $1$ (see \ref{bellman} for details).

The optimization problem from Eq. \ref{opt1} reduces (see \ref{computations}) to
\begin{equation} \label{opt2}
\max_{\nu} \left\{ \nu \left( \frac{\lambda}{\nu + 1} - c \frac{\lambda}{\nu} - C \right) \right\} ~,
\end{equation}
where, moreover,
\begin{eqnarray} \label{E_l}
E_{\nu, \lambda} [\ell] & = & \frac{\lambda}{\nu} ~,
\end{eqnarray}
and
\begin{eqnarray}
E_{\nu, \lambda} [n] & = & \frac{\lambda}{\nu + 1}  \label{E_n} ~.
\end{eqnarray}
Imposing the extremization in Eq. \ref{opt2} leads to
\begin{equation}
\frac{\lambda}{\nu + 1} - \frac{\lambda \nu}{( \nu + 1 )^2} - C = 0 ~.
\end{equation}
Setting $\mu=1$ and re--normalizing the other rates with respect to $\mu$, the optimal processing rate of the queue is
\begin{equation} \label{nu}
\nu^* = \sqrt{\frac{\lambda}{C} } - 1 ~.
\end{equation}

The dependence on the costs $c$ is restored by the condition that the optimal rate of utility is not negative
(since not answering at all would guarantee zero payoffs and costs).
Imposing $U^* \geq 0$,
\begin{eqnarray}
U^* & = \frac{\lambda}{\nu ^* + 1} - c \frac{\lambda}{\nu^*} - C \nonumber \\
~~ & = \frac{\lambda}{\sqrt{\frac{\lambda}{C} }} - c \frac{\lambda}{\sqrt{\frac{\lambda}{C} } -1} - C   \geq   0 \nonumber \\
\Longrightarrow & \nu^*  \geq  \frac{\sqrt{c}}{1 - \sqrt{c} } ~,
\end{eqnarray}
where $\frac{\sqrt{c}}{1 - \sqrt{c} }$ is always non negative for $c \leq 1$.
In summary, the two conditions that ensure individual strategy's optimization are
\begin{equation}   \label{equation1}
\nu^* = \left\{
\begin{array}{lll}
\sqrt{\frac{\lambda}{C} } - 1 & \mbox{ if } & \sqrt{\frac{\lambda}{C} } - 1 \geq \frac{\sqrt{c}}{1 - \sqrt{c} } \\
0 & \mbox{ otherwise } & ~.
\end{array}
\right.
\end{equation}

In general, one can safely assume the buffer capacity to be infinite, but for some reasons that will be clear later,
 one can also be interested to the case in which the queue length $L$ is finite.
The maximization problem (Eq. \ref{opt2}) is now (see \ref{computations})
\begin{equation} \label{opt_L}
\max_{\nu} \left\{ \left[ 1 - \left( \frac{\lambda}{\lambda+\nu} \right)^L \right] \left( \frac{\lambda \nu}{\nu + 1} - c \lambda \right) - C \nu \right\} ~.
\end{equation}
The problem in Eq. \ref{opt_L} has a unique solution which depends on both $c$ and $C$.
The solution is however non--algebraic, even in the case $c=0$.

\section{Collective behavior} \label{interaction}

We consider now many agents located on the nodes of a network, with a given topological structure. Individually, each agent behaves as discussed
in the previous section, but now individual $\lambda$'s depend on the collective behavior.
For the sake of simplicity we consider a homogeneous network, of fixed degree $K$, and assume that all agents behave similarly.
Each one of them receives tasks exogenously at a Poisson rate $\lambda_0$ (also normalized with respect to $\mu$).
When an agent processes a task contained in her buffer, this task is forwarded to all neighbors with a constant and uniform
probability $p>0$. This is the probability that a given task is forwarded to a specific neighbor or a probability of successful transmission.
In the framework of e-mail networks, it can be the probability that a friend is interested in the subject of a given e-mail.
Under this scenario,  an agent receives tasks from outside (at a rate $\lambda_0$) and from neighbors (at an unknown rate).
The overall rate $\lambda$ is estimated by the agents themselves under the assumption that the arrival events are Poissonian.
This is a good assumption as far as $p \ll 1$.
Note that, assuming homogeneity, $\lambda$ is the same, in expectation, for all the nodes of the network.

The number of tasks $\ell_i$ arriving in the buffer of agent $i$ is partly coming from her $N_i$ neighbors and partly from outside, in symbols
\begin{equation} \label{total_ell}
\ell_i = \ell_{0\rightarrow i} + \sum_{j \in N_i} \ell_{j \rightarrow i}~,
\end{equation}
and similarly the number of surviving tasks is
\begin{equation}
n_i = n_{0\rightarrow i} + \sum_{j \in N_i} n_{j \rightarrow i}~.
\end{equation}
We will proceed in steps to study the stochastic processes associated to these variables.
The first step is that of deriving coupled equations for the generating functions of this stochastic process on the
network.

\subsection{Generating functions approach to the stationary state}

Since the process is based on the assumption that arrival events are Poisson distributed, the generating functions
formalism \cite{W90} turns out to be very useful to analyze the stationary collective behavior of the system.
Moreover, the procedure (detailed in \ref{computations}) holds for any network structure under local tree approximation.

The overall generating function for the probability that $i$ has received a certain
number of tasks $\ell_i$ (between two consecutive processing events) is defined by
\begin{eqnarray}
E \left[ s^{\ell_i} \right] = \Xi_i (s) & = & \frac{\nu_i}{\nu_i + (1-s) \lambda_0} \prod_{j \in N_i} \frac{\nu_i}{\nu_i + \left[ 1 - \Psi_j \left( 1 + (s-1)p \right) \right] \nu_j } ~. \label{final_ell}
\end{eqnarray}
The overall generating function $\Psi_i(s)$ for the number $n_i$ of tasks that are still active when $i$ process the buffer is instead
\begin{equation} \label{final_n}
E \left[ s^{n_i} \right] = \Psi_i (s) = \frac{\nu_i + 1}{\nu_i + 1 + (1-s) \lambda_0} \prod_{j \in N_i} \frac{\nu_i}{\nu_i + \left[ 1 - \Psi_j \left( 1 + (s-1)p \frac{\nu_i}{\nu_i +1} \right) \right] \nu_j } ~ .
\end{equation}

\subsection{Stationary state of the regular network}

Eq. \ref{final_n} can be easily solved in a regular network of degree $K$.
In this case, we expect $\nu_i = \nu_j$ for any $i$ and $j$, and Eqs. \ref{final_ell}, \ref{final_n} become
\begin{equation} \label{system}
\left\{
\begin{array}{rcl}
\Xi (s) & = & \frac{\nu}{\nu + (1-s) \lambda_0} \left( \frac{1}{ 2 - \Psi \left( 1 + (s-1)p \right) } \right)^K \\
\Psi (s) & = & \frac{\nu + 1}{\nu + 1 + (1-s) \lambda_0} \left( \frac{1}{2  - \Psi \left[ 1 + (s-1)p \frac{\nu}{\nu +1}  \right] \nu } \right)^K
\end{array}
\right.
\end{equation}
The expected values for $\ell_i$ and $n_i$ can be computed as the derivatives of generating functions in $s=1$.
From Eq. \ref{system} we get
\begin{equation} \label{E_n1}
E [n] = \Psi' (1) = \frac{\lambda_0}{1 + (1 - pK) \nu} ~.
\end{equation}
From
\begin{equation}
\Xi' (1) = \frac{\lambda_0}{\nu} + p K \Psi' (1)
\end{equation}
we get
\begin{equation} \label{E_ell}
E [ \ell ] = \Xi' (1) =  \frac{\lambda_0 (\nu+1)}{\nu (1 + (1-pK)\nu) } ~.
\end{equation}
In Eqs. \ref{E_n1} and \ref{E_ell} the relevant parameters are $\lambda_0$, $\nu$ and the product $pK$, which represents the expected number of forwarded tasks per processed one by a single node.
We will call it
\begin{equation}
m \equiv pK ~.
\end{equation}

Noting that the expected rate $\lambda$ is actually $ \nu E [ \ell ] $, we have that
\begin{equation} \label{lambda}
\lambda = \frac{\lambda_0 (\nu+1)}{1 + (1-m)\nu } ~.
\end{equation}
Using Eq. \ref{lambda} and the optimality relation $\nu = \sqrt{\frac{\lambda}{C} } - 1$,
we obtain the condition
\begin{equation}   \label{equation2}
\frac{\lambda_0}{C} = (1-m) \nu^2 + (2-m) \nu +1 ~,
\end{equation}
from which the optimal processing rate $\nu$ can be determined.

\subsection{Approach to the stationary state: stability under a learning process}

Note that the condition in Eq. \ref{equation2} determines the stationary state of the queueing process on the network.
Starting from a random initial condition, i.e. random distribution of processing rates $\{\nu_{i}\}$,
the system should self-organize to reach some stationary state satisfying the above relation.
A realistic way to generate this self-organization is by means of a learning process for the quantity $\lambda_i$ for each node $i$ of the system.
Assume that agents update their expectations about $\lambda_i$ every time they process the buffer (times are indexed by discrete $t$'s), using the linear rule
\begin{equation} \label{learning}
\lambda_i (t+1) = (1-\epsilon) \lambda_i (t) + \epsilon \frac{\ell_i (t)}{T_i (t)} ~,
\end{equation}
where $\ell_i (t)$ is the number of tasks arrived in $i$'s buffer, and $T_i (t)$ is the actual time--lag between $t$ and $t+1$.

In expectations, Eq.\ref{learning} becomes trivially
\begin{equation}
\lambda_i (t+1) = \lambda_i (t) + \epsilon \Big( \nu_i E [ \ell_i ] - \lambda_i(t) \Big) ~ .
\end{equation}
If we substitute $\epsilon t' = t$, going at the limit $\epsilon \rightarrow 0$ we obtain the dynamical equation
\begin{equation} \label{dyn_eq}
\frac{d \lambda_i}{d t'} =  \nu_i E_{\nu_i} [ \ell_i ] - \lambda_i (t')  ~,
\end{equation}
where $\nu_i = \sqrt{\frac{\lambda_i}{C}} - 1$ is a function of $\lambda_i$, and $E_{\nu_i} [ \ell_i ]$ is given by Eq. \ref{E_ell}.

Imagine a small perturbation around the stationary value $\lambda_i^*$, so that $\lambda_i(t') = \lambda_i^* + \delta \lambda_i (t')$.
Eq. \ref{dyn_eq} gives
\begin{eqnarray}
\frac{d \delta \lambda_i}{d t'} & = & \lambda_i^* + \frac{\partial }{\partial \nu_i} \left( \nu_i E_{\nu_i} [ \ell_i ] \right) \frac{\partial \nu_i}{\partial \lambda_i}
\delta \lambda_i - \lambda_i^* - \delta \lambda_i + O (\delta \lambda_i^2) \nonumber \\
& = & \left( \frac{\lambda_0 m}{[1+ (1-m)\nu_i]^2} \cdot \frac{1}{2 \sqrt{\lambda_i C} }
 - 1 \right) \delta \lambda_i + O (\delta \lambda_i^2) ~.
\end{eqnarray}
The system is stable if the term in the brackets is negative, i.e.
\begin{equation}
\frac{\lambda_0 m}{[1+ (1-m)\nu_i]^2} \cdot \frac{1}{2 \sqrt{\lambda_i C} } - 1 \leq 0 ~,
\end{equation}
substituting with Eq. \ref{lambda},
\begin{equation}
\frac{\lambda_i^2 m}{\lambda_0 (\nu_i+1)^2} \cdot \frac{1}{2 \sqrt{\lambda_i C} }  \leq 1 ~.
\end{equation}
Finally, using Eq. \ref{nu}, we get the stability condition
\begin{equation} \label{stab_cond}
\frac{\lambda_0}{C}  \geq  \frac{m}{2} \nu_i + \frac{m}{2} ~.
\end{equation}
that is always satisfied under Eq. \ref{equation2}, when $m \leq 1$.
When instead $m>1$, Eq. \ref{equation2} is a downward parabola and Eq. \ref{stab_cond} is satisfied only as long as $\frac{\lambda_0}{C}$ is
increasing in $\nu_i$ (see Section \ref{results}).

Note that, as we defined the learning process, agents update their beliefs only when processing their buffer.
This means in principle that they could assume $E [ \lambda_i ]$ to be actually smaller than $\lambda_0$,
because they don't know $\lambda_0$. In this case they could all play the strategy $\nu_i=0$ and never update beliefs.
The rate $\lambda_0$ could be very high but agents would never find out.
To avoid similar paradoxical situations, in the following, we will assume that agents know $\lambda_0$ and also know that $E [ \lambda_i ] \geq \lambda_0$.
In other terms, Eq.\ref{learning} becomes
\begin{equation}
\lambda_i (t+1) = \max \Big\{ (1-\epsilon) \lambda_i (t) + \epsilon \frac{\ell_i(t)}{T(t)} \ , \lambda_0 \Big\} ~ .
\end{equation}
As expected from the regularity of the network, in the stable case all the $\nu_i$s will converge to the same $\nu$ given by Eq. \ref{equation2}.

The next section and \ref{app_m} are devoted to a detailed analysis of the stationary
collective behavior of the system depending on the external parameters.

\section{Results} \label{results}

The system's stationary behavior depends on Eqs. \ref{equation1} and \ref{equation2}, and
on the stability analysis of the learning process (Eq. \ref{stab_cond}).
The overall scenario emerging from these equations is very rich, since the collective behavior changes
varying $m=pK$,
$\lambda_{0}/C$, and $c$. These form the minimal set of independent external parameters of the model.

We analyze the different behaviors of the system studying the `phase diagram' of $\nu$ as a function of $\lambda_{0}/C$,
when the value of $m$ is fixed with respect to $c$.
As $m$ varies, there are 4 possible distinct cases. We treat here the two most illustrative ones, leaving the remaining two to \ref{app_m}.

\begin{figure}[h]
\begin{center}
\includegraphics[height=8cm,width=12cm]{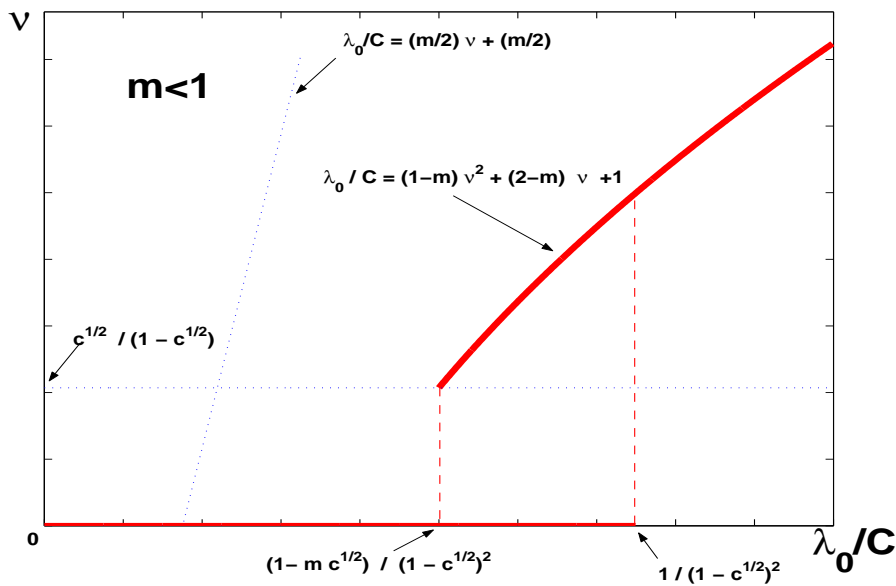}
\end{center}
\caption{When $m=pK<1$ there is at least a stable solution for every value of $\frac{\lambda_0}{C}$, but there are two in the illustrated interval.}
\label{m1}
\end{figure}

\subsection{Case $m \leq 1$} \label{m_leq_1}

When $m \leq 1$, the number of tasks present in the system does not grow, the queue lengths are
bounded, and the dynamics always reach a well defined stationary state. This is the most realistic scenario, illustrated in
Fig. \ref{m1}.
The trivial state in which all agents prefer not to process the tasks, $\nu=0$, is always a possible stable equilibrium as long as $\lambda_{0}/C < \frac{1}{(1-\sqrt{c})^2}$. Above this value, the trivial strategy is suboptimal for any single individual even if adopted by all the others.
Eq. \ref{equation2} predicts the appearance of a non trivial solution with $\nu>0$ for  $\lambda_{0}/C \geq 1$, but the corresponding value of the utility is non-negative only  for $\lambda_{0}/C \geq \frac{1 - m \sqrt{c}}{(1-\sqrt{c})^2}$. The stability condition for this solution (see Eq. \ref{stab_cond}) is always satisfied.

From Figure \ref{m1}, we see that an interval exists, where both solutions are possible and stable.
When $\lambda_{0}/C$ is in this interval, the system will converge to one of the two possible equilibria depending on the initial conditions, and hysteresis-like phenomena can be observed. Note that this co-existence of equilibria is not reflected at the level of single instance: due to the hypothesis of network homogeneity, in a single realization of the dynamics all agents (possibly with different initial conditions) converge to the same state, with $\lambda$ given by Eq.~\ref{lambda}.
The variable cost $c$ does not affect the shape of the functions describing the two possible equilibria, but it affects their region of admissibility and the interval of co-existence. This interval shrinks to zero in $\lambda_0 / C =1$, at the limit $c \to 0$. As $c$ grows this interval is right--shifted and its size grows.

These solutions have been verified numerically using continuous time simulations.

\subsection{Case $m > \frac{2}{1+\sqrt{c}}$}

\begin{figure}[h]
\begin{center}
\includegraphics[height=8cm,width=12cm]{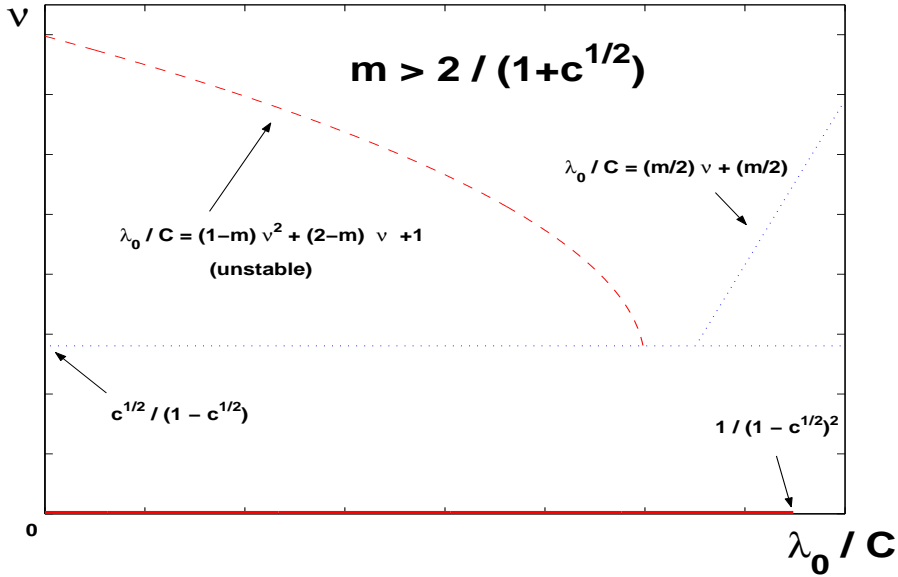}
\end{center}
\caption{When $m=pK>\frac{2}{1+\sqrt{c}}$ the only stable solution is the trivial one, for $\frac{\lambda_0}{C} \leq \frac{1}{\left(1-\sqrt{c}\right)^2}$.}
\label{fig2}
\end{figure}

In this case, the non--trivial solution $\nu>0$ is always unstable, as shown in Fig.~\ref{fig2} obtained combining
Eq.~\ref{equation2} with the conditions imposed by Eqs.~\ref{equation1} and \ref{stab_cond}.
The trivial solution $\nu=0$ exists as long as $\frac{\lambda_0}{C} \leq \frac{1}{(1-\sqrt{c})^2}$.
Above this value,  no stationary solution is expected. The variables $\lambda$ and $\nu$, as well as the buffer's length, will explode to infinity for all nodes, since the generation rate $\lambda_0$ will be higher than the deletion rate (due to $p$, $K$ and the $\nu$'s).
Since  $m>1$, this can happen also for  $\frac{\lambda_0}{C} \leq \frac{1}{(1-\sqrt{c})^2}$, if the initial processing rates are sufficiently
high.
In this case one can speak about \emph{congestion} of the system.
One possibility to avoid congestion is to put a bound $L < \infty$ to the buffers' length: tasks arriving when the buffer is full are neglected.
The corresponding optimization problem (Eq. \ref{opt_L})  can be solved numerically, even if  the optimal value of $\nu$ for $L \gg 1$ will not differ considerably from the previously considered one.
Imposing the bound  $L < \infty$ in the learning process, we get
\begin{equation} \label{learning_L}
\lambda_i (t+1) = (1-\epsilon) \lambda_i (t) + \epsilon \frac{ \min \{ \ell_i (t) , L \} }{T_i (t)} ~.
\end{equation}
Now, implementing this learning process in simulations, we find an additional stationary solution at a large value $\hat{\nu}$, which is independent on the value of the parameter $\frac{\lambda_0}{C}$ (and also on the learning parameter $\epsilon$, necessary for the simulations).

\section{Non--homogeneous networks} \label{non_hom}

It is possible to derive the behavior of the system also in a non--homogeneous network.
Here we restrict to some considerations that generalize previous sections.
Assuming that the structure of the network is uncorrelated, Eq. \ref{final_n} for a $k$-degree node becomes
\begin{equation}
\Psi_k (s) = \frac{\nu_k + 1}{\nu_k + 1 + (1-s) \lambda_0} \left[ \sum_{k'} \frac{k' P(k')}{\left< k \right>}
\frac{\nu_k}{\nu_k + \left[ 1 - \Psi_{k'} \left( 1 + (s-1)p_k \frac{\nu_k}{\nu_k +1} \right) \right] \nu_{k'} } \right]^k ~,
\end{equation}
where $P(k)$ is the degree distribution and $\left< k \right>$ is the average connectivity.
We also generalize to a $p_k$ that depends on the degree $k$ of the sender node.
We can compute
 \begin{equation} \label{non_hom_n}
E [n_k] = \Psi'_k (1) = \frac{\lambda_0}{\nu_k +1} + k \frac{p_k}{\nu_k +1}
\sum_{k'} \frac{k' P(k')}{\left< k \right>} \nu_{k'} \Psi'_{k'} (1) ~,
\end{equation}
from which, calling $\Theta \equiv \sum_{k'} \frac{k' P(k')}{\left< k \right>} \nu_{k'} \Psi'_{k'} (1)$, and using Eqs. \ref{E_l} and \ref{E_n}, we obtain
 \begin{equation}
\lambda_k = \lambda_0 + k p_k \Theta ~.
\end{equation}
The rate $\lambda_k$ at which a node of degree $k$ receives tasks is linear in $k$.
By Eq. \ref{equation1}, giving the optimal $\nu_k$, and using again Eqs. \ref{E_l} and \ref{E_n}, we have
 \begin{equation}
\Theta = \sum_{k p_k > d (\Theta) } \frac{k P(k)}{\left< k \right>}
\left( \lambda_0 + k p_k \Theta \right) \left( 1- \sqrt{ \frac{C}{\lambda_0 + k p_k \Theta} } \right) \equiv F(\Theta) ~,
\end{equation}
where
 \begin{equation}
d(\Theta) = \max \left\{ \frac{ \frac{C}{(1-\sqrt{c})^2} - \lambda_0 }{\Theta}     , \ 0 \right\}
\end{equation}
indicates the minimum $k p_k$ for which it is profitable to check the buffer (i.e. $\nu_k =0$ for $ k p_k \leq d(\Theta)$).

We are looking for the fixed points of $F(\Theta)$, which are steady states of the system.
The behavior is qualitatively the same as in the regular case.
We can distinguish two cases depending on the parameter
 \begin{equation}
m \equiv \frac{\left<  k^2 p_k \right>}{\left< k \right>}  ~,
\end{equation}
which generalizes the $m$ defined in previous sections.

In the case $m>1$, since $F(\Theta) \simeq m \Theta$, for $\Theta \gg 1$, one fixed point diverges to infinity, indicating a congested state.
For $\lambda_0 < \frac{C}{(1-\sqrt{c})^2}$ there is another fixed point in $\Theta=0$, which correspond to the trivial solution $\nu_k=0$ for any $k$.
We are not excluding in principle other fixed points (\ref{app_m} shows how this applies to the regular network case).

Let us now address the case $m \leq 1$. 
When $\lambda_0 \geq \frac{C}{(1-\sqrt{c})^2}$, since $d (\Theta) =0$, it is easy to see that $F(0)>0$, $F'(\Theta) \geq 0$ and $F''(\Theta) \geq 0$ for any $\Theta \geq 0$.
There will be a unique non--trivial fixed point $\Theta^*$ in which all the nodes have a positive processing rate $\nu_k$, for any degree $k$.

Finally, for $m \leq 1$ and $\lambda_0 < \frac{C}{(1-\sqrt{c})^2}$, one fixed point is the trivial one $\Theta=0$.
This is the generalization of the admissible interval of the trivial strategy in the regular network case (see Fig. \ref{m1}).
There could be other fixed points and hysteresis phenomena, as those described in Section \ref{m_leq_1}, depending on the specification of the network structure ($P(k)$) or of the agents behavior ($p_k$).

\section{Conclusion} \label{conc}

Human task-completion dynamics can be formulated as a non-trivial economic optimization problem,
in which individuals optimize their strategies of interaction with the rest of the system.
We have studied a simple model of optimal queueing in a network of identical agents that receive tasks,
store them in queues and process them following a well defined optimal strategy.

The simplicity of the model allows to understand the different roles played by fixed and variable costs ($C$ and $c$ respectively) in the strategy optimization, and more generally it unveils the collective optimal behavior.
Fixed costs $C$ are associated with the action of controlling the state of the queue and processing the tasks, while variable costs $c$ are associated to every received task. This can be interpreted as a way of introducing a cost function that is monotonously increasing with the number of task to process. It is simple to find examples in realistic situations. For instance, in e--mail networks, individuals do not check their mail--boxes all the time as it has a cost in terms of wasted time; hence  they decide to check it at a certain rate that they consider optimal.

We have shown that varying fixed costs does not change the type of transition from the not responding equilibrium to the dynamical steady states, but the position of the transition itself (fixing $\lambda_0$, individuals are favored to respond as $C$ decreases).
The dependence on the variable costs is instead more surprising:  when variable costs are absent, the transition is continuous, whereas for $c>0$ a discontinuity is developed that introduces a non-trivial hysteresis phenomenon in the collective behavior of the processing rate $\nu$.

This model is a first step towards the understanding of the effects of individual optimization processes in collective human dynamics.
Its simplicity allows to obtain several analytical results and a complete theoretical insight in the problem. There are many possible extensions of the model, it could be possible to introduce a multiple-queue system with some priority principle, as proposed in Ref. \cite{KC03} for a similar problem.

Another possibility could be to consider heterogeneous networks of agents, pushing forward the analysis of Section \ref{non_hom}.
As discussed there, we expect the general behavior of an uncorrelated network to qualitatively resemble the regular network case.
It would be interesting to go beyond random networks, including e.g. degree correlation, and to
study the network structure that realizes the best compromise between stability of the system, average utility and minimization of lost information.
We leave this analysis for future investigations.

\appendix

\section{Bellman equation} \label{bellman}

As in dynamic programming we compute the optimal strategy $\nu^*$ using the recursive Bellman equation for the value function $V$ with given initial condition $\nu_0$ and with time-discount rate $\delta = e^{-\beta}$, i.e.
\begin{equation} \label{vibetaopt}
 V_{\beta}^{*}(\nu_0 ,\lambda) = \max_{\nu}\left\{\mathbf{E}_{\nu , \lambda}\left[ \left(1-e^{-\beta}\right) U(\nu , \lambda) + e^{- \beta  T_{0}} V_{\beta}^{*}(\nu_1 , \lambda) \right]\right\}~,
\end{equation}
where $T_{0}$ is the time at which the queue is processed for the first time, the constant factor $1-e^{-\beta}$ is applied to every utility outcome (its role will be clarified in the following) and $\mathbf{E}_{\nu, \lambda}[\cdot]$ is the average value of a quantity over the distribution of the arrival and processing Poisson processes.
As the process is assumed to be stationary, $T_{j}=T$ for all $j$ and the equation simplifies to
\begin{equation} \label{vibeta}
 V_{\beta}^{*}(\nu; \lambda) = \max_{\nu}\left\{\frac{ \left(1-e^{-\beta}\right) \mathbf{E}_{\nu,\lambda} \left[ U(\nu; \lambda)\right]}{1- \mathbf{E}_{\nu} \left[ e^{- \beta  T(\nu)}\right]} \right\}~,
\end{equation}
The factor $1-e^{-\beta}$ allows to take the limit $\beta \to 0$ of the Bellman equation keeping the value function measurable and giving the same weight to all future events.
In this limit, the optimization problem reduces to Eq. \ref{opt1}.

\section{Details of the analytic derivation} \label{computations}

\subsection{Derivation of Eq. \ref{opt2}}

In order to derive an expression for the optimal processing rate (that is also the response rate) of an individual, we have to
compute the expectations in Eq. \ref{opt1}. The probability that the queue contains $\ell$ tasks when it is observed is
\begin{eqnarray} \label{P_l}
P_{\nu, \lambda} (\ell) & = & \int_0^{\infty} \frac{(\lambda t)^{\ell} e^{-\lambda t}}{a!} \nu e^{-\nu t} dt
= \left( \frac{\lambda}{\lambda + \nu} \right)^{\ell} \frac{\nu}{\lambda + \nu} ~,
\end{eqnarray}
from which the expectation value
\begin{eqnarray} \label{B_E_l}
E_{\nu, \lambda} [\ell] & = & \sum_{\ell=0}^{\infty} \ell P_{\nu, \lambda} (\ell) = \frac{\lambda}{\nu} ~.
\end{eqnarray}
The conditional probability that only $n$ tasks survive (i.e. have non-zero payoff when the queue is checked) out of the $\ell$ arrived ones reads
\begin{eqnarray}
P_{\nu, \lambda, \mu} (n | \ell) & = & {\ell \choose n} \left( \frac{\lambda}{\nu + \mu} \right)^n \left( \frac{\mu}{ \nu + \mu} \right)^{\ell-n} ~.
\end{eqnarray}
Remember that, without loss of generality, we assume $\mu=1$ (and the relative normalization of $\nu$ and $\lambda$).
We obtain the expectation value of $n(\nu, \lambda)$,
\begin{eqnarray}
E_{\nu, \lambda} [n] & = & \sum_{\ell=0}^{\infty} P_{\nu, \lambda} (\ell) \sum_{n=0}^{\ell} n P_{\nu, \lambda} (n| \ell) \nonumber \\
& = & \sum_{\ell=0}^{\infty} \left( \frac{\lambda}{\lambda + \nu} \right)^{\ell} \frac{\nu}{\lambda + \nu} \sum_{n=0}^{\ell} n {\ell \choose n} \left( \frac{\lambda}{\nu + 1} \right)^n \left( \frac{1}{ \nu + 1} \right)^{a-n} \nonumber \\
& = & \sum_{\ell=0}^{\infty} \left( \frac{\lambda}{\lambda + \nu} \right)^{\ell} \frac{\nu}{\lambda + \nu} \cdot \frac{ \ell \nu}{\nu + 1}
= \frac{\nu}{\nu + 1} \cdot \frac{\lambda}{\nu} = \frac{\lambda}{\nu + 1}  \label{B_E_n} ~.
\end{eqnarray}
The optimization problem from Eq. \ref{opt1} reduces to
\begin{equation}
\max_{\nu} \left\{ \nu \left( \frac{\lambda}{\nu + 1} - c \frac{\lambda}{\nu} - C \right) \right\} ~.
\end{equation}

\subsection{Derivation of Eq. \ref{opt_L}}

Eq. \ref{opt1} is correct for $\ell < L$, but now $P_{\nu,\lambda} (\ell )=0$ for every $\ell > L$, hence
\[ P_{\nu,\lambda} (L ) = 1 - \sum_{\ell=0}^{L-1} P_{\nu,\lambda} (\ell ) ~. \]
The expected value of $\ell$ is computed in the same spirit of Eq. \ref{B_E_l}, but differently,
\begin{eqnarray} \label{E_l_L}
E_{\nu, \lambda} [\ell]
& = & \sum_{\ell=0}^{L-1} \ell P_{\nu, \lambda} (\ell) + L \left( 1 - \sum_{\ell=0}^{L-1} P_{\nu,\lambda} (\ell ) \right)
= \frac{\lambda}{\nu} \left[ 1 - \left( \frac{\lambda}{\lambda+\nu} \right)^L \right] ~.
\end{eqnarray}
Similarly Eq. \ref{B_E_n} becomes
\begin{equation} \label{E_n_L}
E_{\nu, \lambda} [n] =  \frac{\lambda}{\nu + 1} \left[ 1 - \left( \frac{\lambda}{\lambda+\nu} \right)^L \right] ~ ,
\end{equation}
and the maximization problem (Eq. \ref{opt2}) is now
\begin{equation}
\max_{\nu} \left\{ \left[ 1 - \left( \frac{\lambda}{\lambda+\nu} \right)^L \right] \left( \frac{\lambda \nu}{\nu + 1} - c \lambda \right) - C \nu \right\} ~.
\end{equation}

\subsection{Derivation of Eqs. \ref{final_ell} and \ref{final_n}}

The number of tasks arrived at node $i$ from outside the system between two successive accesses to the queue has distribution
\begin{equation}
P ( \ell_{0 \rightarrow i} = k ) = \left( \frac{\lambda_0}{\lambda_0 + \nu_i} \right)^k \frac{\nu_i}{\lambda_0 + \nu_i} ~,
\end{equation}
and the corresponding generating function reads
\begin{equation} \label{ell_0}
\psi_{0 \rightarrow i} (s) = \sum_{k=0}^{\infty} s^k P ( \ell_{0 \rightarrow i} = k ) = \frac{\nu_i}{\nu_i + (1-s) \lambda_0}  ~.
\end{equation}
We have called $\nu_{i}$ the optimal processing rate of the queue at node $i$.
The probability distribution of the number of tasks with non-zero payoff is
\begin{equation}
P ( n_{0 \rightarrow i} = k ) = \sum_{a=0}^{\infty} P (a) P (k|a) = \frac{1+\nu_i}{\lambda_0 + \nu_i + 1} \left( \frac{\lambda_0}{\lambda_0 + \nu_i + 1} \right)^k ,
\end{equation}
and the corresponding generating function is
\begin{equation} \label{n_0}
\phi_{0 \rightarrow i} (s) = \frac{\nu_i + 1}{\nu_i + 1 + (1-s) \lambda_0}  ~.
\end{equation}

The same computations for $\ell_{j \rightarrow i}$ and $n_{j \rightarrow i}$ are more complicated.
Starting from the relation
\begin{equation}
\ell_{j \rightarrow i} = \sum_{h=0}^{m_j} \ell_{j \rightarrow i}^{(h)}
\end{equation}
we first have to compute the probability that $j$ discharges $m_j$  times her buffer before $i$ discharges her own,
i.e.
\begin{equation} \label{first_step}
P (m_j = k) = \left( \frac{\nu_j}{\nu_i + \nu_j} \right)^k \frac{\nu_i}{\nu_i + \nu_j}   ~.
\end{equation}
The generating function of this process is
\begin{equation}
\xi_{j \rightarrow i} (z) = \sum_{k=0}^{\infty} \left( \frac{\nu_j}{\nu_i + \nu_j} \right)^k \frac{\nu_i}{\nu_i + \nu_j} z^k
= \frac{\nu_i}{\nu_i + (1-z) \nu_j} ~.
\end{equation}
In each of these discharges the number of tasks forwarded to $i$ has distribution
\begin{equation}
P (\ell^{(h)}_{j \rightarrow i} = k) = \sum_{n=k}^{\infty} P_j (n) {n \choose k} p^k (1-p)^{n-k}  ~,
\end{equation}
where $P_j (n)$ is the probability that $j$ finds $n$ still active tasks when checking (i.e. with positive payoffs).
The generating function for $P (\ell^{(h)}_{j \rightarrow i})$  can be expressed in terms of the generating
function for $P_j(n)$,
\begin{eqnarray}
\Phi_{j \rightarrow i} (s) & = & \sum_{k=0}^{\infty} P (\ell^{(h)}_{j \rightarrow i} = k) s^k =
\sum_{k=0}^{\infty} \sum_{n=k}^{\infty} P_j (n) {n \choose k} (ps)^k (1-p)^{n-k} \nonumber \\
& = & \sum_{n=k}^{\infty} \sum_{k=0}^{\infty}  P_j (n) {n \choose k} (ps)^k (1-p)^{n-k} =
\sum_{n=k}^{\infty} P_j (n) \left( 1 + (s-1)p \right)^{n}  \nonumber \\
& \equiv & \Psi_j \left( 1 + (s-1)p \right) ~,
\end{eqnarray}
so that $\Psi_j$ is the generating function of $j$'s buffer length.

Putting things together, the generating function for the distribution of the number of tasks that $i$ receives from $j$ is
\begin{equation} \label{last_step}
\xi_{j \rightarrow i} \left( \Phi_{j \rightarrow i} (s) \right) = \frac{\nu_i}{\nu_i + \left[ 1 - \Psi_j \left( 1 + (s-1)p \right) \right] \nu_j } ~ .
\end{equation}
Recalling Eqs. \ref{total_ell} and \ref{ell_0}, the overall generating function for the probability that $i$ has received a certain
number of tasks $\ell_i$ (between two consecutive processing events) is defined by
\begin{eqnarray}
\Xi_i (s) & = & \psi_{0 \rightarrow i} (s) \prod_{j \in N_i}   \xi_{j \rightarrow i} \left( \Phi_{j \rightarrow i} (s) \right) \nonumber \\
& = & \frac{\nu_i}{\nu_i + (1-s) \lambda_0} \prod_{j \in N_i} \frac{\nu_i}{\nu_i + \left[ 1 - \Psi_j \left( 1 + (s-1)p \right) \right] \nu_j } ~.
\end{eqnarray}

One can repeat the whole procedure of Eqs. \ref{first_step}--\ref{last_step}, from Eq. \ref{n_0}, to obtain the overall generating function $\Psi_i(s)$ for the number $n_i$ of tasks that are still active when $i$ process the buffer,
\begin{equation}
\Psi_i (s) = \frac{\nu_i + 1}{\nu_i + 1 + (1-s) \lambda_0} \prod_{j \in N_i} \frac{\nu_i}{\nu_i + \left[ 1 - \Psi_j \left( 1 + (s-1)p \frac{\nu_i}{\nu_i +1} \right) \right] \nu_j } ~ .
\end{equation}

\section{Other results} \label{app_m}

We continue here, from Section \ref{results}, the study of the phase diagram of $\nu$ as a function of $\lambda_{0}/C$, as $m$ varies with respect to $c$.
When $m>1$, congestion is always a possible outcome of the system, but it may coexist with stable non--trivial steady states, as long as $m \leq \frac{2}{1 + \sqrt{c}}$.
Since $c<1$ there is an interval for $m$ that characterize two possibilities, analyzed in this Appendix.

Let us start by defining $m' \equiv 2 \frac{1-\sqrt{2 \sqrt{c} - c}}{(1-\sqrt{c})^2}$.
Note that, for $c \in (0,1)$, we always have
$ 1 < m' < \frac{2}{1 + \sqrt{c}} $.

\subsection*{Case $1 < m < m'$}

\begin{figure}[h]
\begin{center}
\includegraphics[height=8cm,width=12cm]{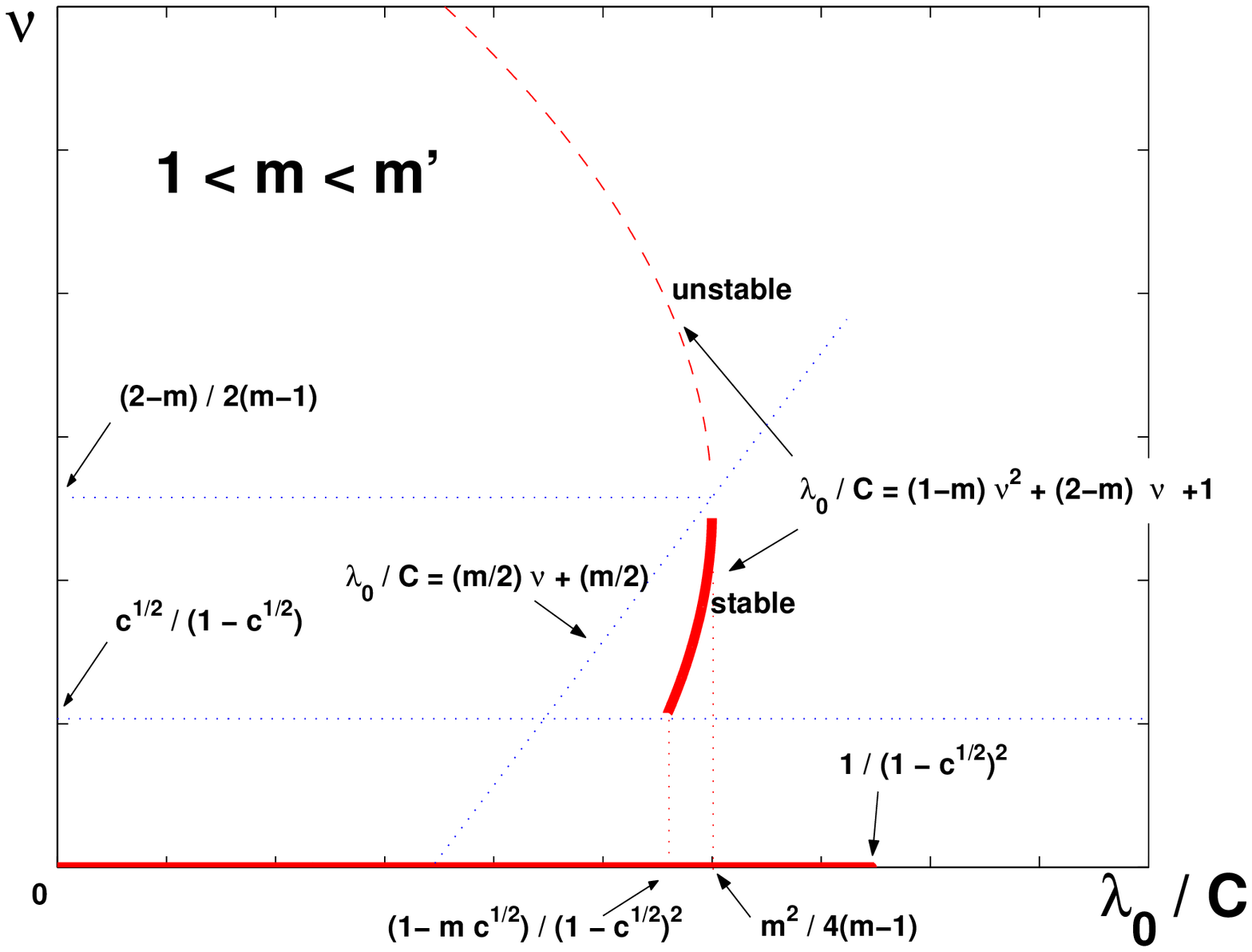}
\end{center}
\caption{When $m=pK>1$ there are no stable solution for $\frac{1}{(1 - \sqrt{c})^2}$, but there are two in the illustrated interval.}
\label{m2}
\end{figure}

When $m>1$ the parabola determined by Eq. \ref{equation2} is leftward.
We have shown that, when $m > \frac{2}{1 + \sqrt{c}} >1$, the only stable solution is the trivial one ($\nu=0$), as long as $\frac{\lambda_0}{C} < \frac{1}{(1-\sqrt{c})^2}$.
The case $m \in \left( 1, m' \right)$, illustrated in Fig. \ref{m2}, is more interesting.
The stability condition (Eq. \ref{stab_cond}), intersects Eq. \ref{equation2} at the vertex of the parabola, so that the upper branch of the solution is always unstable.
This point is given by $\frac{\lambda_0}{C} = \frac{m^2}{4(m-1)} $ and $\nu=\frac{2-m}{2 (m-1)}$.
Note that when $m >1$, then
\begin{equation} \label{ineq}
m < m'  \ \ \Rightarrow \ \ \frac{m^2}{4(m-1)} < \frac{1}{(1-\sqrt{c})^2} ~.
\end{equation}
If $\frac{1 - m \sqrt{c}}{(1-\sqrt{c})^2} < \frac{\lambda_0}{c} < \frac{m^2}{4(m-1)} $ there is another stable solution (see Fig. \ref{m2}) which lies on the parabola.
There is an interval with two stable non--intersecting equilibria, but one of them is valid only inside this interval, while the other is on a proper (at both sides) overset of it.
We assume not to observe any phenomenon of hysteresis, since agents will keep adopting the trivial strategy, no matter how we vary the parameter $\frac{\lambda_0}{c}$ across the interval $ \left[ \frac{1 - m \sqrt{c}}{(1-\sqrt{c})^2} , \frac{m^2}{4(m-1)} \right] $.

\subsection*{Case $m' \leq m \leq \frac{2}{1 + \sqrt{c}}$}

If $m' \leq m \leq \frac{2}{1 + \sqrt{c}}$, then Eq. \ref{ineq} is not satisfied and there is hysteresis, as  for the case $m\leq 1$.
For $\frac{\lambda_0}{C} < \frac{1}{(1-\sqrt{c})^2}$ the trivial solution is the only stable one.
For $\frac{\lambda_0}{C} \in \left[ \frac{1}{(1-\sqrt{c})^2} , \frac{1 - m \sqrt{c}}{(1-\sqrt{c})^2} \right]$ we have two stable equilibria (the trivial and the non--trivial one).
Finally, for $\frac{\lambda_0}{C} > \frac{1 - m \sqrt{c}}{(1-\sqrt{c})^2} $, only the non--trivial one is possible, up to $ \frac{\lambda_0}{C} =\frac{m^2}{4(m-1)}$.
Cases $m=m'$ and $m=\frac{2}{1 + \sqrt{c}}$ represent the same situation at the limit of zero measures.

\end{document}